\documentclass{IET-Conf-Paper}
\usepackage{graphicx}  
\usepackage{subfigure}    
\usepackage{enumerate}
\counterwithout{figure}{section}
\counterwithout{equation}{section}
\counterwithout{table}{section}

\begin{document}

\title{GRIDLESS TOMOGRAPHIC SAR IMAGING BASED ON ACCELERATED ATOMIC NORM MINIMIZATION WITH EFFICIENCY}

\author{Silin Gao\ad{1,2,3}, Zhe Zhang\ad{1,4,5}\corr, Bingchen Zhang\ad{1,2,3}, Yirong Wu\ad{1,2,3}}

\address{
    \add{1}{Aerospace Information Research Institute, Chinese Academy of Sciences, Beijing 100094, China}
    \add{2}{Key Laboratory of Technology in Geo-spatial Information Processing and Application System, Chinese Academy of Sciences, Beijing 100190, China}
    \add{3}{School of Electronic, Electrical and Communication Engineering, University of Chinese Academy of Sciences, Beijing 100049, China}
    \add{4}{Key Laboratory of Intelligent Aerospace Big Data Application Technology, Suzhou 215123, China}
    \add{5}{Suzhou Aerospace Information Research Institute, Suzhou 215123, China}
    \email{zhangzhe01@aircas.ac.cn}}


\begin{abstract}
    Synthetic aperture radar (SAR) tomography (TomoSAR) enables the reconstruction and three-dimensional (3D) localization of targets based on multiple two-dimensional (2D) observations of the same scene. The resolving along the elevation direction can be treated as a line spectrum estimation problem. However, traditional super-resolution spectrum estimation algorithms require multiple snapshots and uncorrelated targets. Meanwhile, as the most popular TomoSAR imaging method in modern years, compressed sensing (CS) based methods suffer from the gridding mismatch effect which markedly degrades the imaging performance. As a gridless CS approach, atomic norm minimization can avoid the gridding effect but requires enormous computing resources. Addressing the above issues, this paper proposes an improved fast ANM algorithm to TomoSAR elevation focusing by introducing the IVDST-ANM algorithm, which reduces the huge computational complexity of the conventional time-consuming semi-positive definite programming (SDP) by the iterative Vandermonde decomposition and shrinkage-thresholding (IVDST) approach, and retains the benefits of ANM in terms of gridless imaging and single snapshot recovery. We conducted experiments using simulated data to evaluate the performance of the proposed method, and reconstruction results of an urban area from the SARMV3D-Imaging 1.0 dataset are also presented.
\end{abstract}

\maketitle

\section{Introduction}
Synthetic aperture radar (SAR) uses the synthetic aperture principle to achieve high resolution in both the azimuth and range directions. SAR tomography (TomoSAR) technique extends the synthetic aperture principle to the third dimension, the elevation direction, and enables reconstruction and three-dimensional localization of targets. TomoSAR observes the same scene multiple times from various angles of view via either multi-baseline flights or antenna arrays to collect cross-track information and construct the synthetic aperture along the elevation direction~\cite{tomosar}.

The resolving along the elevation direction is a line spectrum estimation problem and can be solved via several methods. Traditional super-resolution spectrum estimation algorithms, such as CAPON~\cite{capon} and MUSIC~\cite{music}, do not suffer from Rayleigh resolution limit but require multiple snapshots, and in practice the temporal averaging is usually replaced by the spatial averaging which is not a good estimation in some cases~\cite{spectrum}. In recent years, compressed sensing (CS) technique~\cite{cs2006,cs2007,cs2008} has been introduced into TomoSAR, considering the scene is usually highly sparse along the elevation direction~\cite{l1}. Compared with subspace algorithms, CS algorithms are believed to be capable of achieving super-resolution with few samples and require only a single snapshot. As a discrete domain model, CS algorithms usually divide the scene into virtual grids and find the sparest support over the grid via greedy methods or convex optimizations. Obviously, this is an approximation, and in real applications the targets are usually not on the grid precisely, resulting in the ``gridding mismatch effect'' which takes in the form of ``energy leak'' at neighboring grids and harms the sparsity.

In recent years, a gridless CS approach, namely atomic norm minimization (ANM)~\cite{anm} emerged. ANM avoids the gridding effect by allowing the targets to be continuously distributed in the scene and reconstructing the target scene directly by semi-positive definite programming (SDP), without utilizing the virtual grid. ANM also works under a single snapshot scenario. TomoSAR is usually highly under-sampled, which strengthened the significance of the gridding effect. This inspired us to introduce ANM into TomoSAR to achieve a much more robust three-dimensional imaging with insufficient samples.

However, the price of ANM is the huge computational complexity brought by SDP, which limits the application of ANM and makes it infeasible in practice. Tackling this issue, a straightforward approach is to replace the time-consuming SDP with simple gradient descent and shrinkage, named as iterative Vandermonde decomposition and shrinkage-thresholding (IVDST) as introduced in~\cite{ivdst}. IVDST can reduce the computational complexity from $\mathcal{O}( N^{3.5} )$ to $\mathcal{O}( N^{2} )$ without losing of reconstruction accuracy, where $N$ is the length of samples.

This paper introduces the ANM algorithm using IVDST (IVDST-ANM) to TomoSAR for elevation focusing. The rest of this paper is organized as follows. Section \ref{section:model} presents the imaging model of TomoSAR and recalls the fundamentals of ANM. The proposed TomoSAR approach with ANM is introduced in Section 3, with the acceleration of IVDST-ANM. The potential and limitations of proposed method are discussed in Section 4 with simulated data. We applied the proposed method to real data collected from the urban area of Yuncheng, Shanxi Province, China, and show its effectiveness in Section 5, followed by the conclusions in Section 6.

\section{Model and Problem Statement}\label{section:model}

\subsection{TomoSAR Signal Model}
\begin{figure}[h]\vspace*{-1em}
    \centering
    \includegraphics[width=0.7\columnwidth]{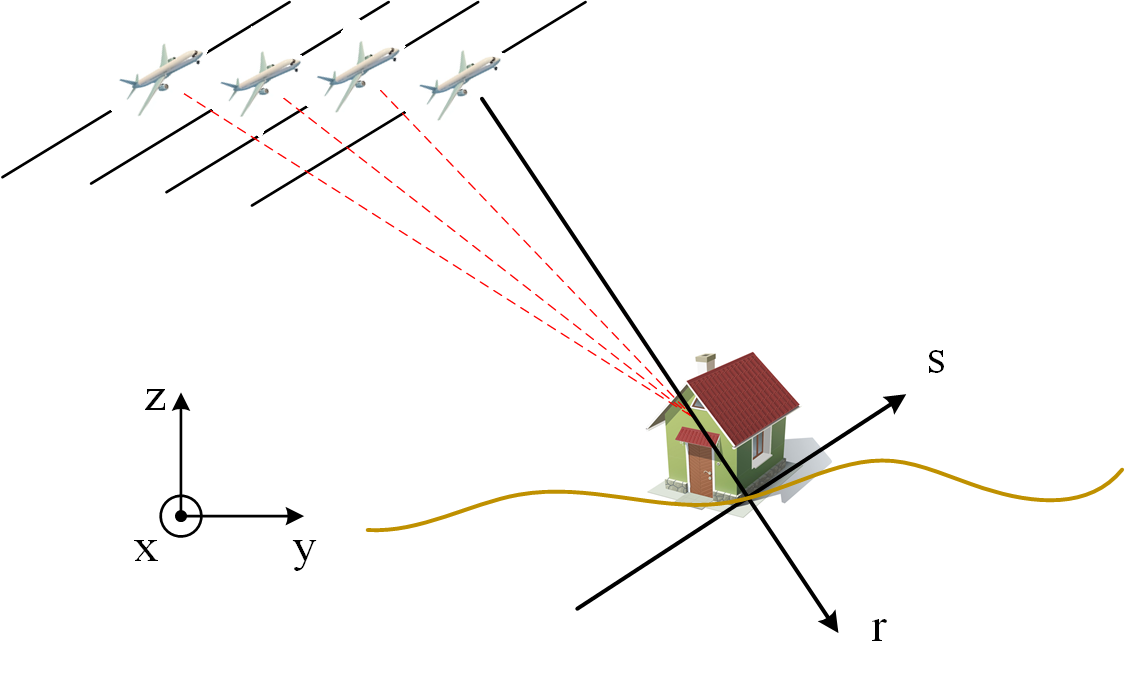}\vspace*{-1em}
    \caption{TomoSAR imaging model diagram.}
    \label{fig1}\vspace*{-1em}
\end{figure}

The traditional SAR imaging system obtains two-dimensional images in azimuth and range, which are projections of real three-dimensional targets on a two-dimensional plane. Targets located at different heights on the same azimuth-range pixel will have their imaging results superimposed, resulting in overlapping. TomoSAR imaging model is shown in Fig. \ref{fig1}.

A TomoSAR imaging systems is usually a multi-channel system, with each channel being either a fight pass or an antenna array element. SLCs achieved by each channel are aligned and stacked into a tensor. Denote the SLC image of the $n-$th channel as $\boldsymbol{g}_n$, then the signal located at azimuth-range pixel $( x_0, r_0 )$ can be expressed as:
\begin{equation}\label{tomosar}
    \boldsymbol{g}_n( x_0,r_0 ) =\int{\gamma}( x_0,r_0,s ) \exp \left( j\frac{4\pi}{\lambda}\cdot \frac{sb_n}{r_0} \right) \text{d}s+\boldsymbol{\omega }( x_0,r_0 ) ,
\end{equation}
where $s$ is the elevation axis, $b_n$ is the baseline length the $n-$th channel and $\boldsymbol{\omega }$ is the additive Gaussian white noise.

By estimating the spatial spectrum of $\boldsymbol{g}_n( x_0,r_0 )$, the resolving along the elevation direction can be obtained. Theoretically, the Rayleigh resolution in the elevation direction is
\begin{equation}
    \rho =\frac{\lambda r}{2\Delta s},
\end{equation}
where $\Delta s$ is the array aperture along the elevation direction.

\subsection{Compressed sensing (CS) and Atomic Norm Minimization (ANM)}
\begin{figure}[h]\vspace*{-1em}
    \centering
    \includegraphics[width=0.8\columnwidth]{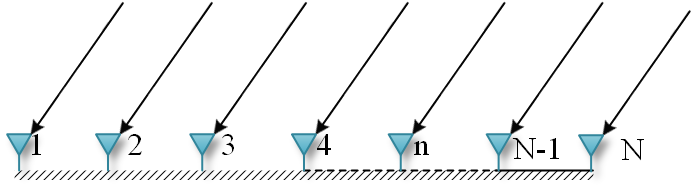}\vspace*{-1em}
    \caption{Array model diagram.}
    \label{fig2}\vspace*{-1em}
\end{figure}

For simplicity, we only consider the elevation direction of TomoSAR imaging and omit the range-azimuth processing. Usually, a TomoSAR imaging system achieves SLCs by antenna array elements along the elevation direction, so that the imaging model can be treated as an array imaging problem. In this paper, we only consider the array as a uniform linear array (ULA) with $N$ elements and there are $K$ targets located at different angles of views, as shown in Fig. \ref{fig2}.

The array manifold of the $k-$ th target can be written as
\begin{equation}
    \boldsymbol{a}( f_k ) =\left( \begin{array}{c}
            1                                \\
            e^{j2\pi f_k}                    \\
            \vdots                           \\
            e^{j2\pi ( N-1 ) f_k} \\
        \end{array} \right) ,
\end{equation}
where $f_k\in ( 0,1 )$ is the normalized digital frequency of the $k-$th target. For noise free case, the echo sum of all $K$ targets is
\begin{equation}
    \boldsymbol{x}=\sum_k{c_k}\boldsymbol{a}\left( f_k \right) ,
\end{equation}
where $c_k$ is the reflectivity of the $k-$th target.
In order to locate the targets, we need to recover  $\left\{ f_k \right\}$ (and subsequentially $\left\{ c_k \right\}$) from the observation $\boldsymbol{x}$. In canonical compressed sensing algorithms, we discretize $f$ into $M$ virtual grids and define a basis set of steering vectors $\boldsymbol{a}( f )$  as $\mathcal{A}_M=\left\{ \boldsymbol{a}( f ) :\,\,f\in \left( 0,\frac{1}{M},\frac{2}{M},\cdots ,\frac{M-1}{M} \right) \right\}$.

Then, the sparsity of target scene can be defined as the $\ell _0$ norm of $\boldsymbol{x}$ over the basis $\mathcal{A}_M$ as:
\begin{equation}\label{csl0}
    \lVert \boldsymbol{x} \rVert _{\mathcal{A}_M,0}=\text{inf}\left\{ K:\,\,\sum_{k=1}^K{\hat{c}_k\boldsymbol{a}\left( \hat{f}_k \right) =\boldsymbol{x}},\,\,\boldsymbol{a}\left( \hat{f}_k \right) \in \mathcal{A}_M \right\} .
\end{equation}
We can then recover $\left\{ f_k \right\}$ and $\left\{ c_k \right\}$ via popular CS methods by minimizing the sparsity defined in (\ref{csl0}).

Usually the targets are off-grid, i.e., $\boldsymbol{a}\text{(}f_k\text{)}\notin \mathcal{A}_M$, resulting in the gridding mismatch issue. ANM solves this issue by manipulating $M\rightarrow \infty$, leveraging $\mathcal{A}=\left\{ \boldsymbol{a}\left( f \right) :\,\,f\in ( 0,1 ) \right\} $.
Then, we can define the gridless version of (\ref{csl0}) as
\begin{equation}
    \lVert \boldsymbol{x} \rVert _{\mathcal{A},0}=\text{inf}\left\{ K:\,\,\sum_{k=1}^K{\hat{c}_k\boldsymbol{a}\left( \hat{f}_k \right) =\boldsymbol{x,\,\,a}\left( \hat{f}_k \right) \in \mathcal{A}} \right\} ,
\end{equation}
as well as its convexified version by relaxation
\begin{equation}\label{anm}
	\begin{split}
		&\lVert \boldsymbol{x} \rVert _{\mathcal{A}} =\lVert \boldsymbol{x} \rVert_{\mathcal{A},1}                                                                                                                                              \\
		& =\text{inf}\left\{ \sum_k{\left| \hat{c}_k \right|}:\,\,\sum_k^{}{\hat{c}_k\boldsymbol{a}\left( \hat{f}_k \right) =\boldsymbol{x,\,\,a}\left( \hat{f}_k \right) \in \mathcal{A}} \right\} .
	\end{split}
\end{equation}

Theoretically, $\left\{ f_k \right\}$ can be obtained by solving $\lVert \boldsymbol{x} \rVert _{\mathcal{A}}$. Traditionally it is done by solving the following SDP
\begin{equation}\label{sdp}
	\begin{split}
		&\min_{\mathbf{x,}v,\mathbf{T}( \boldsymbol{u} )} v+\frac{1}{N}\text{trace}\left( \mathbf{T}( \boldsymbol{u} ) \right) \\
		&\text{s.t.~~}  \left( \begin{matrix}
			v              & \boldsymbol{x}^{\text{H}}               \\
			\boldsymbol{x} & \mathbf{T}\left( \boldsymbol{u} \right) \\
		\end{matrix} \right) \succeq 0     
	\end{split}
    \begin{array}{ll}
                                \\
    \end{array},
\end{equation}
where $\mathbf{T}( \boldsymbol{u} )$ is a Toeplitz matrix determined by its first row $\boldsymbol{u}$. Eq. (\ref{anm}) is a standard SDP and can be solved via popular convex optimization solvers. $\left\{ f_k \right\}$ are embedded in $\mathbf{T}( \boldsymbol{u} )$ and can be solved via Vandermonde decomposition towards it, and $\left\{ c_k \right\}$ can be determined subsequentially without difficulty.

\section{Proposed Method}
\subsection{TomoSAR approach based on ANM}
In the TomoSAR scene, there are usually only a few discrete targets distributed along the elevation axis for a specific azimuth-range pixel, so (\ref{tomosar}) can be discretised as
\begin{equation}
    \boldsymbol{g}_n( x_0,r_0 ) =\sum_k{
        \gamma ( x_0,r_0,s ) \cdot \exp \left( j\frac{4\pi b_ns_k}{\lambda r_0} \right)}+\boldsymbol{\omega }( x_0,r_0 ) .
\end{equation}

For simplicity, assume the baselines of each channel are evenly distributed (i.e. a ULA is form along the cross-track direction),
\begin{equation}
    b_n=\left( n-1 \right) d_S,
\end{equation}
where $d_s$ is the distance of neighbouring array elements along the elevation direction. Define $f_k=\frac{2s_k\cdot d_s}{\lambda r_0}$ as the spatial frequency of the $k-$th target and assume that there is no phase wrapping, thus the received signal can be written as
\begin{equation}
    \boldsymbol{g}( x_0,r_0 ) =\sum_k{\gamma}( x_0,r_0,s_k ) \boldsymbol{a}\left( f_k \right) +\boldsymbol{\omega }( x_0,r_0 ) .
\end{equation}
This is a standard ANM form and can solved via SDP. The algorithm is given in Table \ref{alg1}.

\begin{table}[h]\vspace*{-1em}
    \caption{Standard ANM algorithm via SDP}\vspace*{-1em}
    \begin{center}
        \begin{tabular}{p{\linewidth}}\toprule
            For each pixel $\left( x_0,r_0 \right)$

            \textbf{Input:} $\boldsymbol{g}\left( x_0,r_0 \right)$, $\{s_n\}$, $d_s$ and $K$.

            \textbf{Compute:}

            1. Solve the SDP problem
            \begin{displaymath}
                \begin{array}{ll}
                    \min_{\boldsymbol{g}\left( x_0,r_0 \right),v,\mathbf{T}\left( \boldsymbol{u} \right)} & v+\frac{1}{N}\text{trace}\left( \mathbf{T}\left( \boldsymbol{u} \right) \right) \\
                    \text{s.t}.                                                                           & \left( \begin{matrix}
                            v                                    & \boldsymbol{g}\left( x_0,r_0 \right)^{\text{H}} \\
                            \boldsymbol{g}\left( x_0,r_0 \right) & \mathbf{T}\left( \boldsymbol{u} \right)         \\
                        \end{matrix} \right) \succeq 0.
                \end{array}
                \nonumber
            \end{displaymath}

            2. Decompose $\mathbf{T}\left( \boldsymbol{u} \right)$ into $\mathbf{T}\left( \boldsymbol{u} \right)=\mathbf{A} \mathbf{C}\mathbf{A}^\mathrm{H}$ by root-MUSIC.

            \textbf{Output:} \{${c_k}$\} and \{${s_k=\frac{\lambda r_0}{2d_s}f_k}$\}.
            \\\botrule
        \end{tabular}
        \label{alg1}
    \end{center}\vspace*{-2em}
\end{table}

\subsection{Compressed sensing (CS) and Atomic Norm Minimization (ANM)}
The computational complexity of SDP is $\mathcal{O}\left( N^{3.5} \right)$ where $N$ is the number of channels. This is a hugh computational load. When $N$ is larger than 32, its is usually inapplicable in practice. Tackling this issue, \cite{ivdst} accelerate SDP by utilizing first order gradient descent and shrinkage instead of solving the SDP via interior point method,  which is named as the iterative Vandermonde decomposition and shrinkage-thresholding (IVDST) algorithm.

IVDST utilizes the Vandermonde structure of the array manifold and the sparsity of the signal as \emph{a priori} knowledge to compute the low-rank Toeplitz matrix $\mathbf{T}\left( \boldsymbol{u} \right)$ by shrinkage-thresholding. In each iteration, IVDST enforces the sparsity by low-rankness of $\mathbf{T}\left( \boldsymbol{u} \right)$ (via SVD and shrinkage-thresholding the small singular values) and the positive semi-definition of the constraint matrix (via SVD and eliminating the negative singular values).

Inspired by IVDST, we combine it with the TomoSAR model and propose the IVDST-ANM TomoSAR algorithm. The proposed algorithm is listed in Table \ref{alg2}. Using IVDST instead of SDP, we can reduce the computational complexity of ANM algorithm to $\mathcal{O}\left( N^2 \right)$.

\begin{table}[h]\vspace*{-1em}
    \caption{IVDST-ANM TomoSAR algorithm}\vspace*{-1em}
    \begin{center}
        \begin{tabular}{p{\linewidth}}\toprule
            For each pixel $\left( x_0,r_0 \right)$

            \textbf{Input:}
            $\boldsymbol{g}\left( x_0,r_0 \right)$, $K$, and sampling matrix $\mathbf{P}$.

            \textbf{Initialization:}
            $\mathbf{y} = \mathbf{P}^{-1}\boldsymbol{g}\left( x_0,r_0 \right)$, $\mathbf{T}=\mathcal{T}(\mathbf{y}\mathbf{y}^H)$, $v=\text{tr}(\mathbf{T})/N$, $\boldsymbol{\theta}=(\mathbf{y},v,\mathbf{T})$

            \textbf{Iteration:}

            1. Set $\bar{\boldsymbol{\theta}_i}=\boldsymbol{\theta}_i+\frac{t_{i-1}-1}{t_i}(\boldsymbol{\theta}_i-\boldsymbol{\theta}_{i-1})$, where $t_i=\frac{1+\sqrt{4t_{i-1}^2+1}}{2}$ and $t_0=1$.

            2. Set $\boldsymbol{\theta}_g=(\mathbf{y}_g,v_g,\mathbf{T}_g)$ by $\mathbf{y}_g=\bar{\mathbf{y}_i}-\delta \mathbf{P}^H(\mathbf{P}\bar{\mathbf{y}_i}-\boldsymbol{g}\left( x_0,r_0 \right))$, $v_g=\bar{v_i}$ and $\mathbf{T}_g=\bar{\mathbf{T}_i}$.

            3. Decompose $\mathbf{T}_g$ into $\mathbf{T}_g=\mathbf{V}\boldsymbol{\Lambda}\mathbf{V}^H$.

            4. Shrink $\boldsymbol{\Lambda}$ to $\tilde{\boldsymbol{\Lambda}}$ and set $\tilde{\mathbf{T}_g}=\mathbf{V}\tilde{\boldsymbol{\Lambda}}\mathbf{V}^H$.

            5. Set $\mathbf{Z}=[\text{tr}(\tilde{\boldsymbol{\Lambda}})\mathbf{y}_g^H; \mathbf{y}_g\tilde{\mathbf{T}_g}]$ and decompose $(\boldsymbol{\Sigma},\mathbf{U})=\text{eigs}(\mathbf{Z},\text{rank}(\tilde{\boldsymbol{\Lambda}})+1)$.

            6. Set $\tilde{\mathbf{Z}}=\mathbf{U}\boldsymbol{\Sigma}\mathbf{U}^H$.

            7. Update $\boldsymbol{\theta}_{i+1}=(\mathbf{y}_{i+1},v_{i+1},\mathbf{T}_{i+1})$ by $\mathbf{y}_{i+1}=\tilde{\mathbf{Z}}_{2:end,1}$, $v_{i+1}=\tilde{\mathbf{Z}}_{1,1}$, and $\mathbf{T}_{i+1}=\tilde{\mathbf{Z}}_{2:\text{end},2:\text{end}}$.

            \textbf{Output:} $\mathbf{T}$.
            \\\botrule
        \end{tabular}
        \label{alg2}
    \end{center}\vspace*{-2em}
\end{table}

\section{Simulations}
\subsection{Performance Analysis}
In this section, the IVDST-ANM algorithm is compared to SDP-ANM, traditional CS algorithms OMP and IST using simulated data.

Firstly, a single-scatterer target simulation experiment is conducted. The simulation parameters are shown in Table \ref{tab1}. Assuming that the radar system works at 9.6 GHz, multiple array elements are arranged along the elevation direction, and the data are acquired by one flight for 3D reconstruction. The array consists of 8 elements with $d_s=0.11$ m, and the normalized spatial frequency of single target is located at 0.5.

\begin{table}[h]\vspace*{-1em}
    \caption{Simulation parameters}\vspace*{-1em}
    \setlength\tabcolsep{17pt}
    \begin{center}
        \begin{tabular}{ll}\toprule
            Parameters                              & Values      \\\midrule
            Carrier frequency                       & 9.6 GHz    \\
            Number of array elements                & 8           \\
            Distance of adjacent array elements     & 0.11 m \\
            Number of targets                       & 1           \\
            Normalized spatial frequency of targets & 0.5         \\
            SNR                                     & 30 dB       \\\botrule
        \end{tabular}
        \label{tab1}
    \end{center}\vspace*{-2em}
\end{table}

\begin{figure}[h]\vspace*{-1em}
    \centering
    \subfigure[]{\includegraphics[width = 0.75\columnwidth]{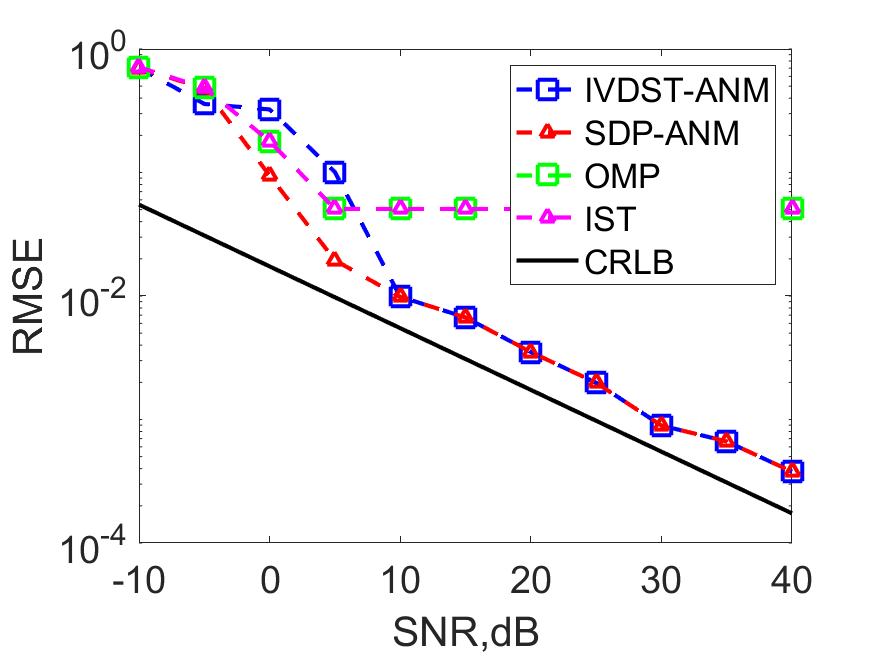}}\vspace*{-0.5em}
    \subfigure[]{\includegraphics[width = 0.75\columnwidth]{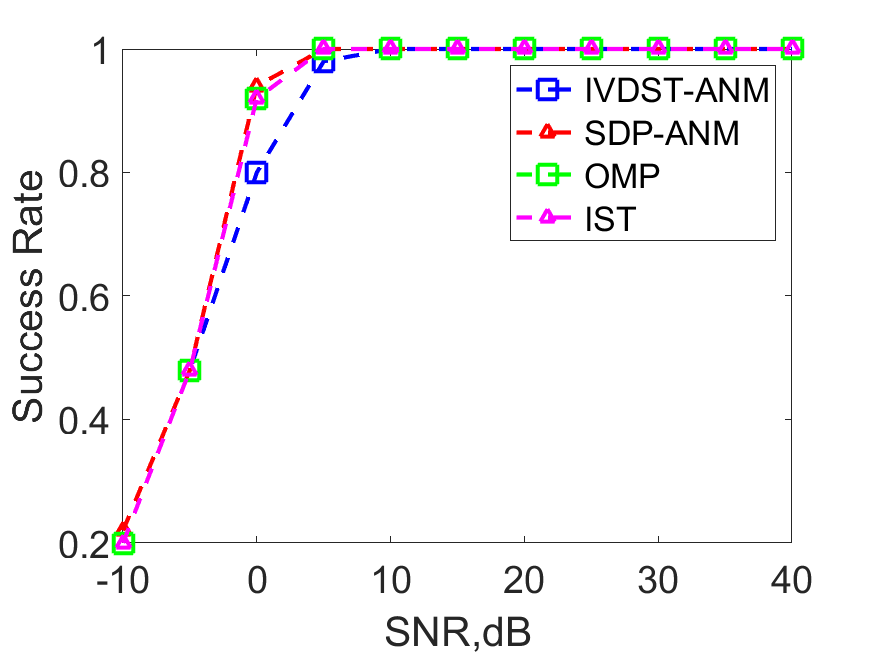}}\vspace*{-1em}
    \caption{Single scatterer estimation quality of IVDST-ANM (blue dashed lines), SDP-ANM (red dashed lines), OMP (green dashed lines), IST (magenta dashed lines) as a function of SNR. (a) RMSE compared to the Cramér Rao Lower Bound (black solid lines), (b) Success rate.}
    \label{snr}\vspace*{-1em}
\end{figure}

We gradually increase SNR from -10 dB to 40 dB and compare the performance of the four algorithms and the results are shown in Fig. \ref{snr}. The RMSEs of the four algorithms are shown in (a) and they are compared with the Cram\'er-Rao lower bound (CRLB). The success rate of them is shown in (b). It is obvious that, the performance of IVDST-ANM and SDP-ANM are quite close, and they are both markedly better than CS approaches thanks to their gridless recovery. When SNR is greater than 10 dB, both ANM based methods are approaching the CRLB.

\begin{figure}[h]\vspace*{-1em}
    \centering
    \includegraphics[width=0.75\columnwidth]{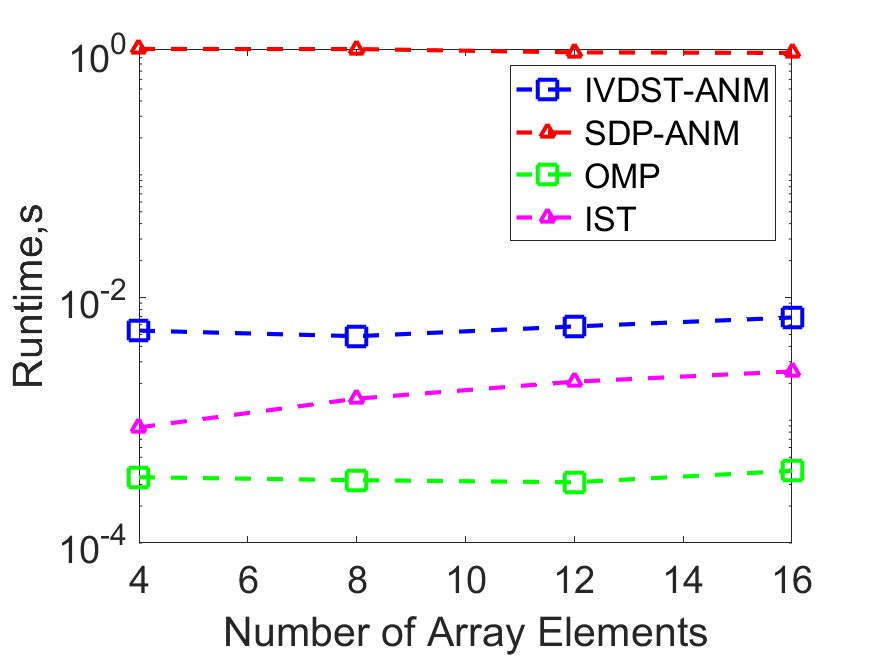}\vspace*{-1em}
    \caption{Runtime of IVDST-ANM (blue dashed lines), SDP-ANM (red dashed lines), OMP (green dashed lines) and IST (magenta dashed lines) as a function of number of array elements.}
    \label{n3}\vspace*{-1em}
\end{figure}

We also tested the running time by varying the number of channels $N$ from 4 to 16. As shown in Fig. \ref{n3}, OMP has the shortest runtime, while IST and IVDST-ANM have a slightly higher runtime and SDP-ANM has a significantly larger runtime than the other three algorithms. Both IST and IVDST-ANM are based on iterative shrinkage-thresholding, so they have similar computational complexity and can keep the runtime between $10^{-3}$ s and $10^{-2}$ s. Therefore, it is a clear advantage to apply IVDST-ANM to TomoSAR. While having almost the same reconstruction accuracy as SDP-ANM, IVDST-ANM substantially reduces the computational complexity.

\begin{figure}[h]\vspace*{-1em}
    \centering
    \includegraphics[width=0.75\columnwidth]{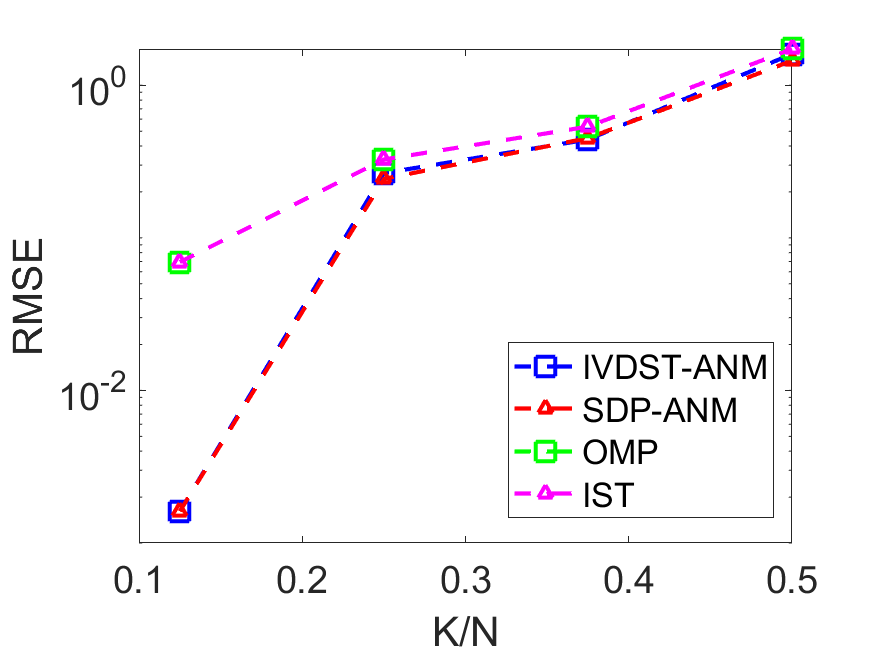}\vspace*{-1em}
    \caption{RMSEs of IVDST-ANM (blue dashed lines), SDP-ANM (red dashed lines), OMP (green dashed lines) and IST (magenta dashed lines) as a function of sparseness.}
    \label{k}\vspace*{-1em}
\end{figure}

We finally vary the sparseness of the scene $K/N$ from 0.1 to 0.5 and compare the performance of the four algorithms, as shown in Fig. \ref{k}. It is obvious that, when the sparseness is below 0.25, both ANM algorithms perform better than the gridded CS algorithms and the performance of IVDST-ANM and SDP-ANM are quite close. For a TomoSAR scene, the sparseness is very low and the overlay times are usually less than 3. Therefore, applying IVDST-ANM into the processing of TomoSAR 3D reconstruction can improve the performance significantly without adding too much computational complexity.

\subsection{TomoSAR Scene Simulation}
\begin{figure}[h]\vspace*{-1em}
    \begin{minipage}[h]{1.0\linewidth}
        \centering
        \subfigure[]{\includegraphics[width = 0.5\linewidth]{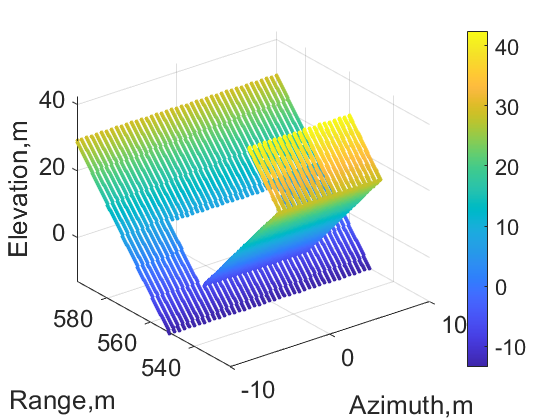}}\hfill
        \subfigure[]{\includegraphics[width = 0.5\linewidth]{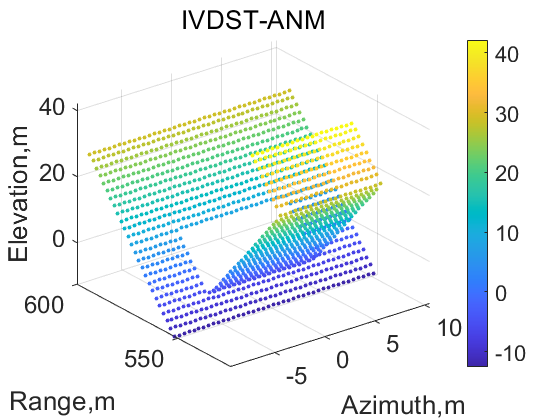}}\vspace*{-0.5em}
        \subfigure[]{\includegraphics[width = 0.5\linewidth]{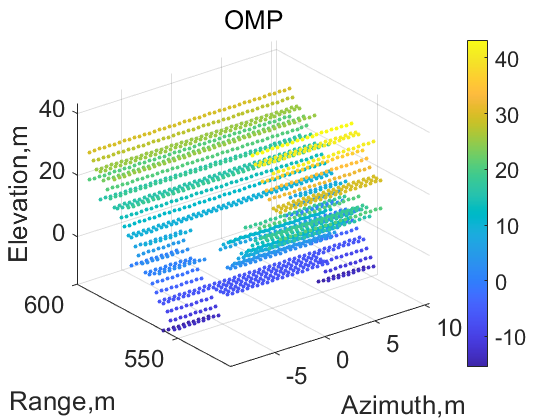}}\hfill
        \subfigure[]{\includegraphics[width = 0.5\linewidth]{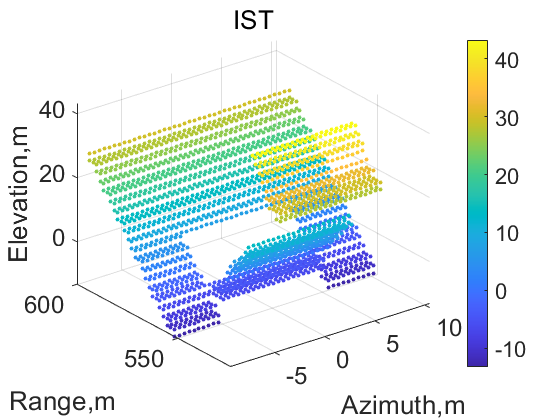}}\vspace*{-1em}
    \end{minipage}
    \caption{Comparison of reconstruction results of simulated data by IVDST-ANM, OMP and IST. (a) Building simulation scenario, (b) IVDST-ANM, (c) OMP, (d) IST.}
    \label{3d}\vspace*{-1em}
\end{figure}

The simulated TomoSAR scene is shown in Fig. \ref{3d}(a), which is a urban area building model. The simulation setup is given in Table \ref{tab2}.

\begin{table}[h]\vspace*{-1em}
    \caption{Simulation parameters}\vspace*{-1em}
    \setlength\tabcolsep{21pt}
    \begin{center}
        \begin{tabular}{ll}\toprule
            Parameters                          & Values       \\\midrule
            Carrier frequency                   & 9.6 GHz     \\
            Number of array elements            & 8            \\
            Distance of adjacent array elements & 0.11 m  \\
            View angle                          & $45^{\circ}$ \\
            Range size                          & 64           \\
            Azimuth size                        & 21           \\
            SNR                                 & 30 dB        \\\botrule
        \end{tabular}
        \label{tab2}
    \end{center}\vspace*{-2em}
\end{table}

The feasibility and superiority of the application of IVDST-ANM in TomoSAR is illustrated by reconstructing it in three dimensions. In this experiment, we use RMSE and runtime as evaluation metrics to measure the performance of the four algorithms.

The results of 3D point clouds reconstruction are shown in Fig. \ref{3d} and evaluation metrics are shown in Table \ref{tab3}. In ground truth Fig. \ref{3d}(a), the walls, roofs and grounds are the main scatterers. Fig. \ref{3d}(b-d) are the point clouds reconstructed by IVDST-ANM, OMP and IST respectively. It can be seen that, benefits from the gridless reconstuction, ANM can achieve better reconstruction for continuous surface targets like walls and roofs, and the reconstruction results are more similar to the ground truth. At the same time, the quality of the 3D point clouds is significantly better than that of gridded CS algorithms.

\begin{table}[h]\vspace*{-1em}
    \caption{Evaluation}\vspace*{-1em}
    \setlength\tabcolsep{19pt}
    \begin{center}
        \begin{tabular}{lll}\toprule
            Algorithm & RMSE                 & Runtime,s \\\midrule
            IVDST-ANM & $7.58\times 10^{-3}$ & 4.45      \\
            SDP-ANM   & $7.35\times 10^{-3}$ & 1928.59   \\
            OMP       & $2.09\times 10^{-1}$ & 1.41      \\
            IST       & $1.13\times 10^{-1}$ & 2.03      \\\botrule
        \end{tabular}
        \label{tab3}
    \end{center}\vspace*{-2em}
\end{table}

\section{Real Data Experiment}
\subsection{Data Set}
\begin{figure}[h]\vspace*{-1em}
    \centering
    \subfigure[]{\includegraphics[width=0.3\textwidth]{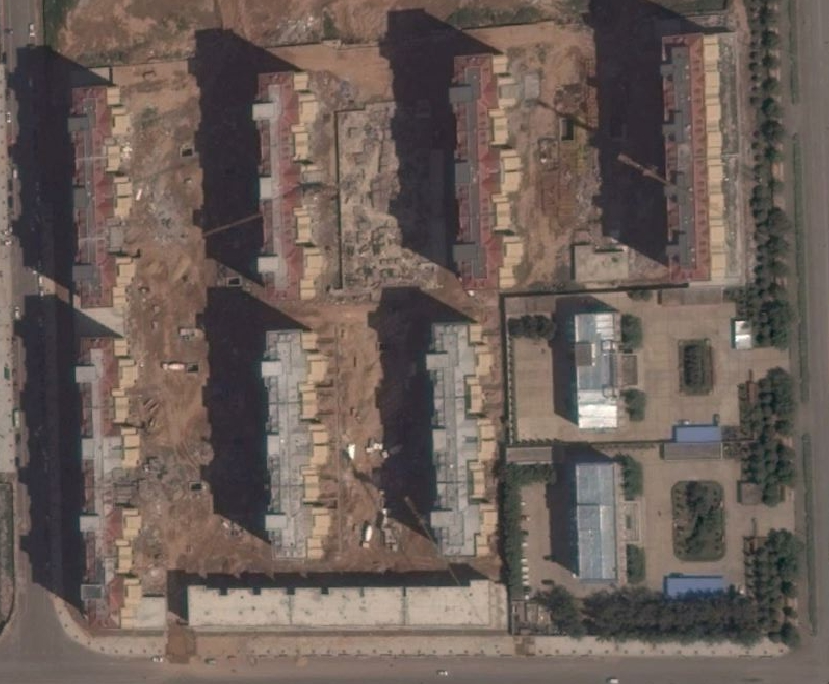}}\vspace*{-0.5em}
    \subfigure[]{\includegraphics[width=0.4\textwidth]{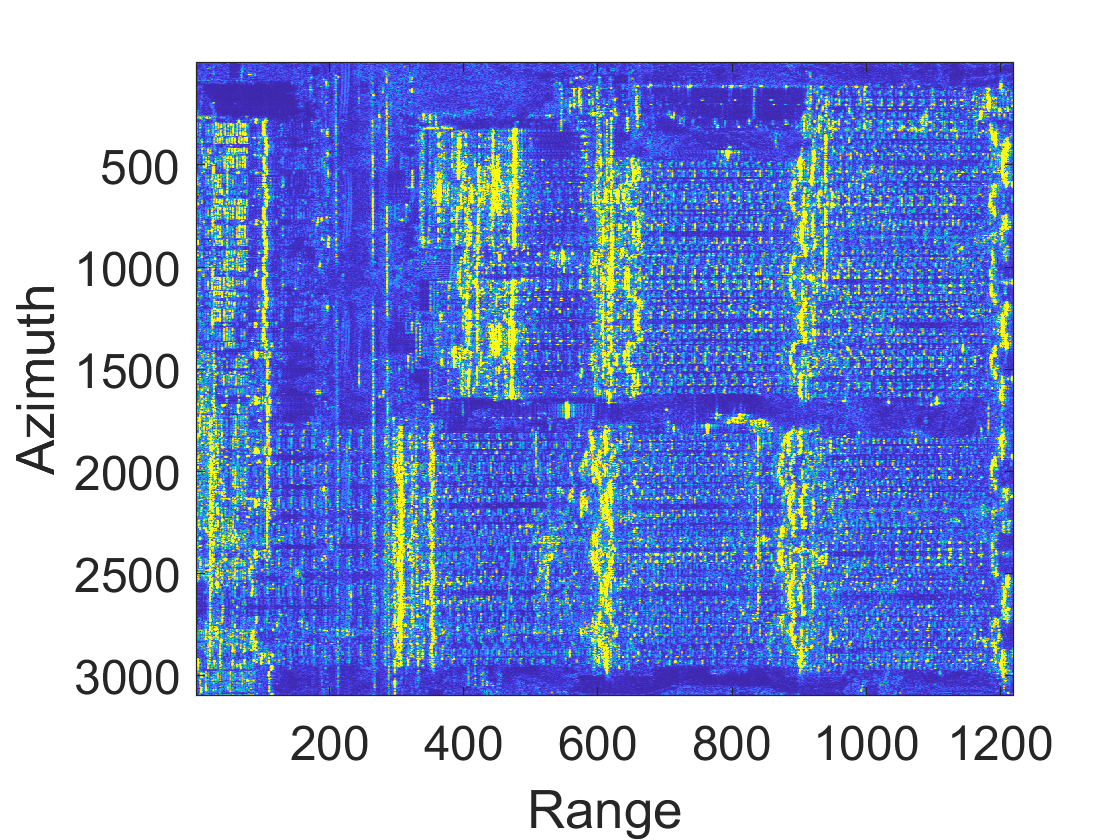}}\vspace*{-1em}
    \caption{Data of Yuncheng area. (a) Optical image~\cite{dataset}, (b) 2D SAR image.}
    \label{optical}\vspace*{-1em}
\end{figure}

We apply the proposed method to SARMV3D-Imaging 1.0 dataset~\cite{dataset}, which is collected by Aerospace and Information Research Institute, Chinese Academy of Sciences. This is an array interferometric SAR, which means that it has a physical cross-track array to achieve the elevation resolution. The antenna equivalent phase centers are distributed along the cross-track direction, and three-dimensional observation data is obtained through a single flight. We choose the data of Yuncheng, Shanxi Province, China, which was observed by this system in 2015. The scene is mainly urban buildings, so the overlay is more serious. The optical image and 2D SAR image of Yuncheng are given in~\cite{dataset}, as shown in Fig. \ref{optical}. The data includes a total of 8 channel images with an image size of 3400 (azimuth) $\times$ 1220 (range), a bandwidth of 500 MHz, and a HH polarization mode.

\subsection{Experiment}
\begin{figure}[h]\vspace*{-1em}
    \centering
    \subfigure[]{\includegraphics[width = 0.4\textwidth]{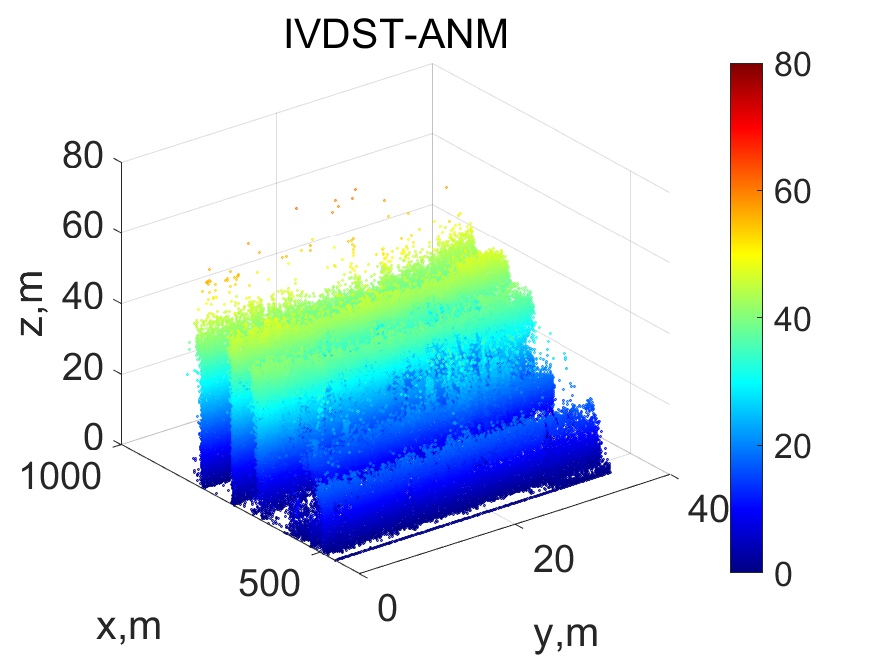}}\vspace*{-0.5em}
    \subfigure[]{\includegraphics[width = 0.4\textwidth]{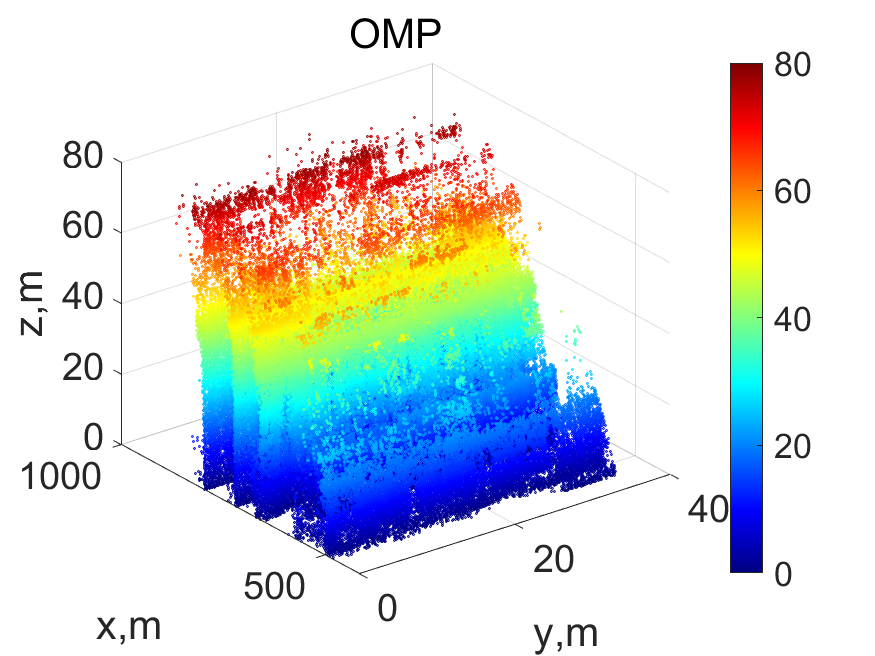}}\vspace*{-0.5em}
    \subfigure[]{\includegraphics[width = 0.4\textwidth]{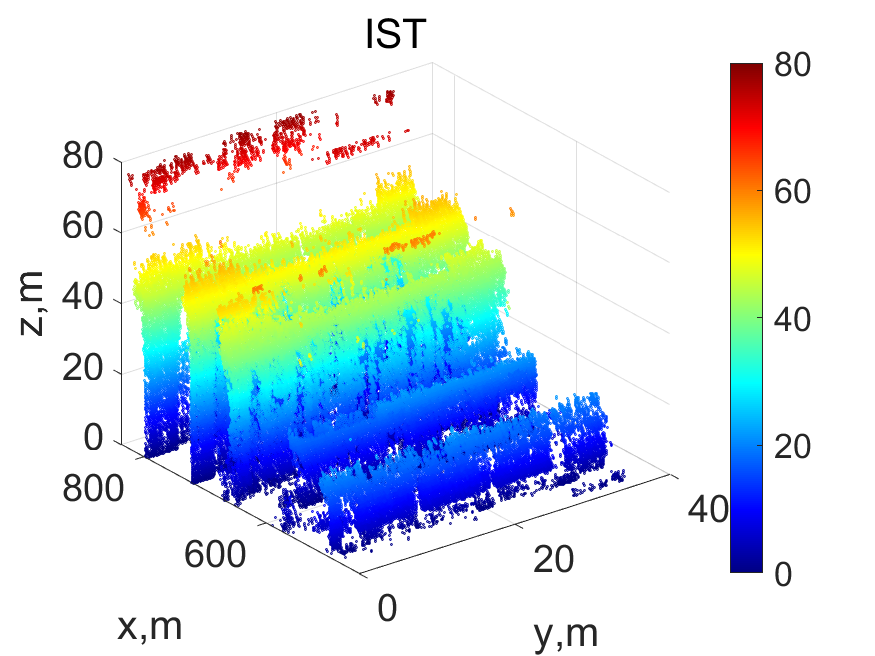}}\vspace*{-1em}
    \caption{3D reconstruction results of real data by IVDST-ANM, OMP and IST . (a) IVDST-ANM, (b) OMP, (c) IST.}
    \label{real}\vspace*{-1em}
\end{figure}

The 3D reconstruction results of the data of Yuncheng are shown in Fig. \ref{real}, where (a), (b) and (c) are the results of IVDST-ANM, OMP and IST respectively. We have already known that the tallest buildings are with height near 50 m. As can be seen in Fig. \ref{real}, the point clouds reconstructed by OMP are the most cluttered, with many fake targets appeared between 50 m and 80 m. The quality of point clouds using IST is better than OMP, but there are still some fake points appeared between 70 m and 80 m. IVDST-ANM reconstructed almost all points between 0 m and 50 m without clutters and fake targets. From the point clouds, the reconstruction result of the proposed method is more consistent with the actual scene and is denser with fewer cluttered points.

\section{Conclusion}

TomoSAR has borad applications and state-of-the-art methods based on CS suffer from gridding effect. Considering that most urban scenes are consisted of a few scattered targets along the elevation direction, ANM is promising gridless super-resolution method. This paper focuses on an improved fast ANM algorithm, and applies the proposed method to 3D height reconstruction of TomoSAR. Compared with the traditional gridded CS algorithms, the proposed algorithm substantially improves the reconstruction performance by avoiding the gridding mismatch effect. Compared with the traditional ANM algorithm, the proposed algorithm greatly accelerates the runtime and satisfies the requirements of practical applications while maintaining almost the same performance. Further work will focus on applying the proposed method to non-uniform baseline acquisitions.

\section{Acknowledgements}

This work is supported by the National Key R\&D Program of China grant \#2018YFA0701903, NSFC grant \#61991421, 61991420 and AIRCAS grant ``Structural sparsity signal high performance adaptive sensing theory and its applications in microwave imaging''.

\end{document}